\documentclass[aps,preprintnumbers,preprint,amsmath,superscriptaddress]{revtex4}
\def\slr#1{\setbox0=\hbox{$#1$}           % set a box for #1
   \dimen0=\wd0                                 % and get its size
   \setbox1=\hbox{/} \dimen1=\wd1               % get size of /
   \ifdim\dimen0>\dimen1                        % #1 is bigger
      \rlap{\hbox to \dimen0{\hfil/\hfil}}      % so center / in box
      #1                                        % and print #1
   \else                                        % / is bigger
      \rlap{\hbox to \dimen1{\hfil$#1$\hfil}}   % so center #1
      /                                         % and print /
   \fi}

\def\ksq{k^2}

\def\mytint#1{\!\int\!\!\frac{d^3\!{#1}}{(2\pi)^3}\,}
\def\gev#1{ GeV${}^{#1}$}
\def\be{\begin{eqnarray}}
\def\ee{\end{eqnarray}}

\usepackage[mathscr]{eucal}
\usepackage{graphicx}

\renewcommand{\theequation}%
    {\arabic{section}.\arabic{equation}}
\makeatletter \@addtoreset{equation}{section} \makeatother

\begin{document}

%\preprint{}

\title{EXCITATIONS OF THE QUARK- GLUON PLASMA}


\author{C. M. Shakin}
\email[]{casbc@cunyvm.cuny.edu}
%\thanks{}
\author{Huangsheng Wang}
\author{Qing Sun}
\author{Hu Li}

\affiliation{%
Department of Physics, Brooklyn College of CUNY\\
Brooklyn, NY 11210, USA}%


\author{Xiangdong Li}
\affiliation{%
Department of Computer System Technology, New York City College of
Technology of CUNY, Brooklyn, NY 11201, USA}%

\date{\today}

\begin{abstract}
We will discuss the spectrum of the eta mesons making use of the
Nambu-Jona-Lasinio (NJL) model supplemented with a model of
confinement. We will go on to discuss the properties of mesons at
finite temperature and the phenomenon of deconfinement. We will
then discuss some excited states of the quark-gluon plasma
calculated in lattice QCD models.These resonances are thought to
be created in heavy-ion collisions.We consider  the role these
states play in leading to a hydrodynamic description of the plasma
at early stages of its formation.
\end{abstract}

\pacs{}
%\keywords{}

\maketitle

\section{Introduction}

In this presentation we will discuss the properties of QCD as one
moves upward along the temperature axis from the point at $T=0$,
where the matter density is zero. Note that for experiments at
RHIC the associated chemical potential is small.

\begin{picture}(430,360)%[htb]
\centering\includegraphics[width=.63\textwidth]{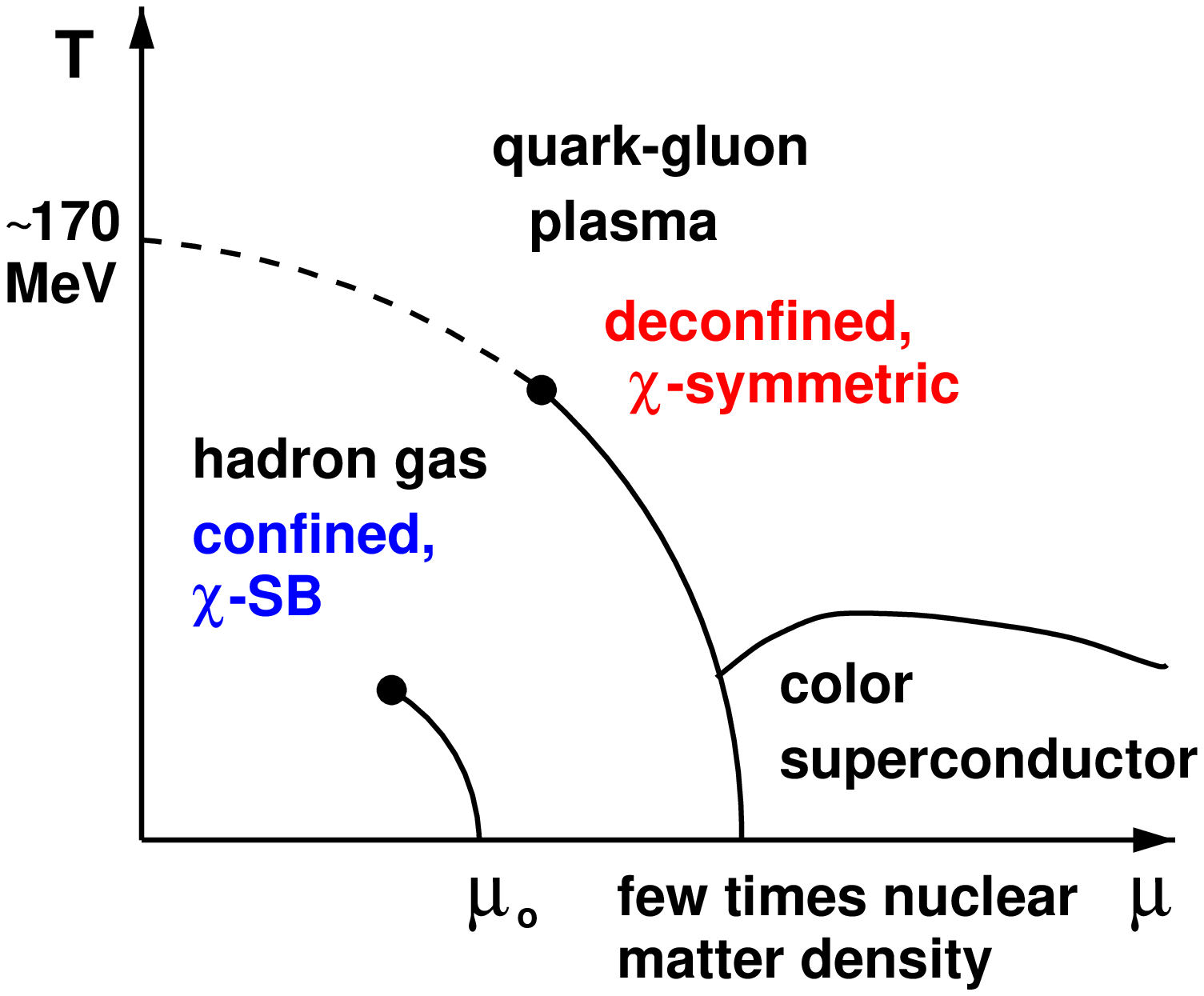}
%\label{aa:phase}
\end{picture}

\textbf{Fig. 1} Schematic phase diagram of nuclear matter (Taken
from F. Karsch, Lect. Notes Phys. \textbf{583}, 209 (2002)
[hep-lat/0106019].)

\section{Properties of the $\eta$ mesons calculated with the NJL interaction and a model of confinement}
\emph{C. M. Shakin and Huangsheng Wang, Physical Review D,
\textbf{65}, 094003 (2002)}\\\\

Recent work has shown that the singlet-octet mixing angles of the
$\eta(547)$ and $\eta^\prime(958)$ are different. That may be
demonstrated either in extended chiral perturbation theory or by
analysis of a large body of experimental data. The conclusion is
that the $\eta(547)$ is almost entirely of octet character, while
the $\eta^\prime(958)$ is mainly of singlet character with about
10\% octet component. It is possible to calculate the mixing
angles and decay constants in our generalized Nambu–-Jona-Lasinio
(NJL) model, which includes a covariant model of confinement. Our
model is able to give a good account of the mass values of the
$\eta(547)$, $\eta^\prime(958)$, $\eta(1295)$, and $\eta(1440)$
mesons. (We also provide predictions for the mass values of a
large number of radially excited states.) It is well known that
the $U_A(1)$ symmetry is broken, so that we only have eight pseudo
Goldstone bosons, rather than the nine we would have otherwise. In
the NJL model that feature may be introduced by including the 't
Hooft interaction in the Lagrangian. That interaction reduces the
energy of the octet state somewhat and significantly increases the
energy of the singlet state, making it possible to fit the mass
values of the $\eta(547)$ and $\eta^\prime(958)$ in the NJL model
when \,\,\,\,\,\,\,\,\,\,\,\,\,\,\,\,\,\,\,\,\,\,\,\,\,\,\,\,\,\,
\,\,\,\,\,\,\,\,\,\,\,\,\,\,\,\,\,\,\,\,\,\,\,\,\,\,\,\,\,\,\,\,\,\,\,\,\,\,\,\,\,\,\,\,\,\,\,\,\,\,\,\,\,\,\,\,\,\,the
't Hooft interaction is included. In this work, we derive the
equations of a covariant random phase approximation that may be
used to study the nonet of pseudoscalar mesons. We demonstrate
that a consistent treatment of the 't Hooft interaction leads to
excellent results for the singlet-octet mixing angles. (The values
obtained for the singlet and octet decay constants are also quite
satisfactory.) It may be seen that the difference between the up
(or down) constituent quark mass and the strange quark mass
induces singlet-octet mixing that is too large. However, the 't
Hooft interaction contains singlet-octet coupling that enters into
the theory with a sign opposite to that of the term arising from
the difference of the quark mass values, leading to quite
satisfactory results. In this work we present the wave function
amplitudes for a number of states of the eta mesons. (The
inclusion of pseudoscalar axial-vector coupling is important for
our analysis and results in the need to specify eight wave
function amplitudes for each state of the eta mesons.) We present
the values of the various constants that parameterize our
generalized NJL model and which give satisfactory values of the
eta meson masses, decay constants, and mixing angles. It is found
that the calculated mass values for the $\eta(1295)$ and
$\eta(1440)$ are quite insensitive to variation of the parameters
of the model whose values have largely been fixed in our earlier
studies of other light mesons.

\begin{picture}(570,600)
 \includegraphics[bb=0 0 571 770, angle=0, scale=0.7]{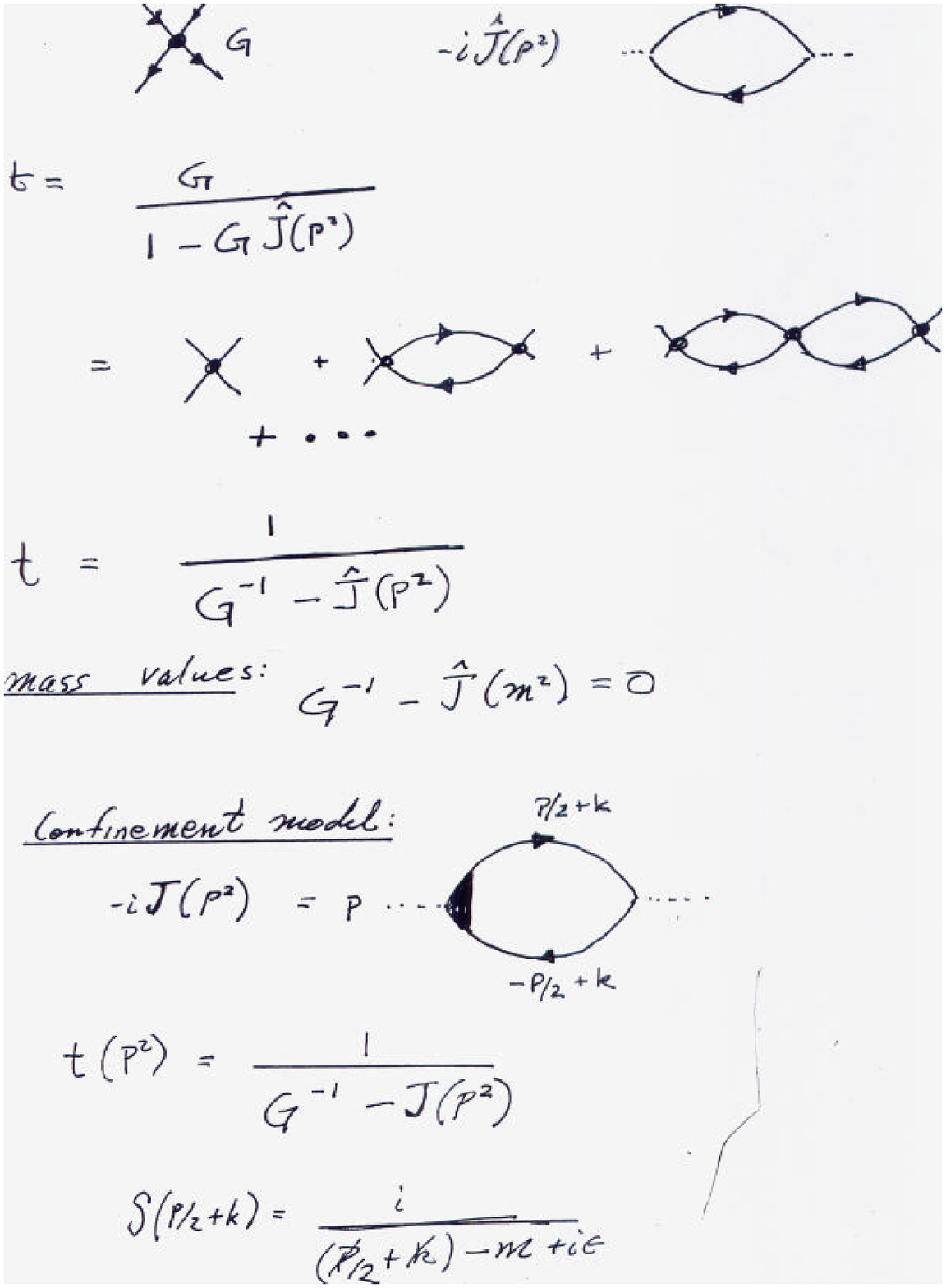}%
% \caption{}
 \end{picture}

 \textbf{Fig. 2}

\begin{picture}(600,500)
 \centering\includegraphics[bb=0 0 600 500, angle=0, scale=0.7]{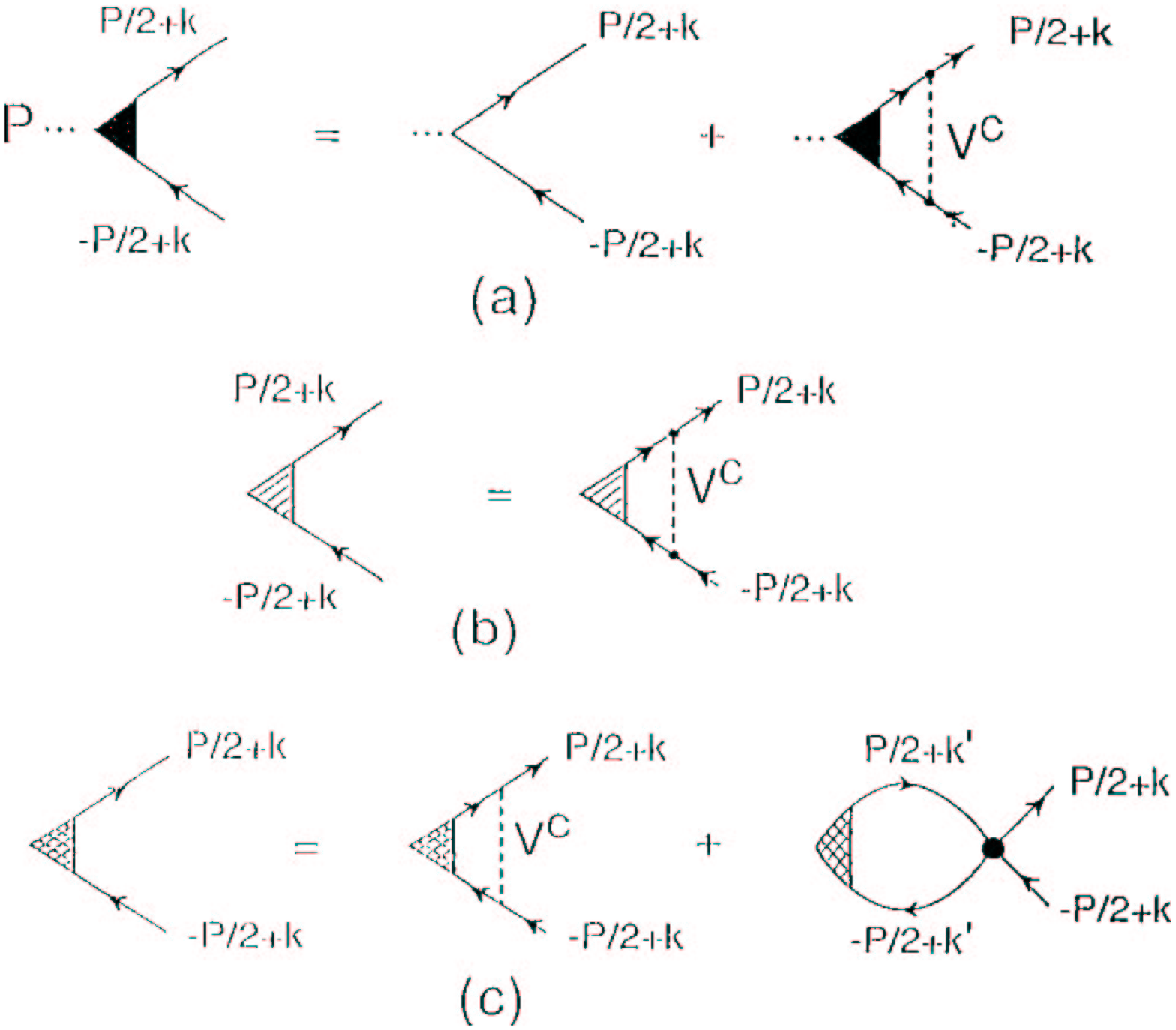}%
 %\caption{}
 \end{picture}

 \textbf{Fig. 3}

 \begin{picture}(600,500)
 \centering\includegraphics[bb=0 0 600 800, angle=0, scale=0.6]{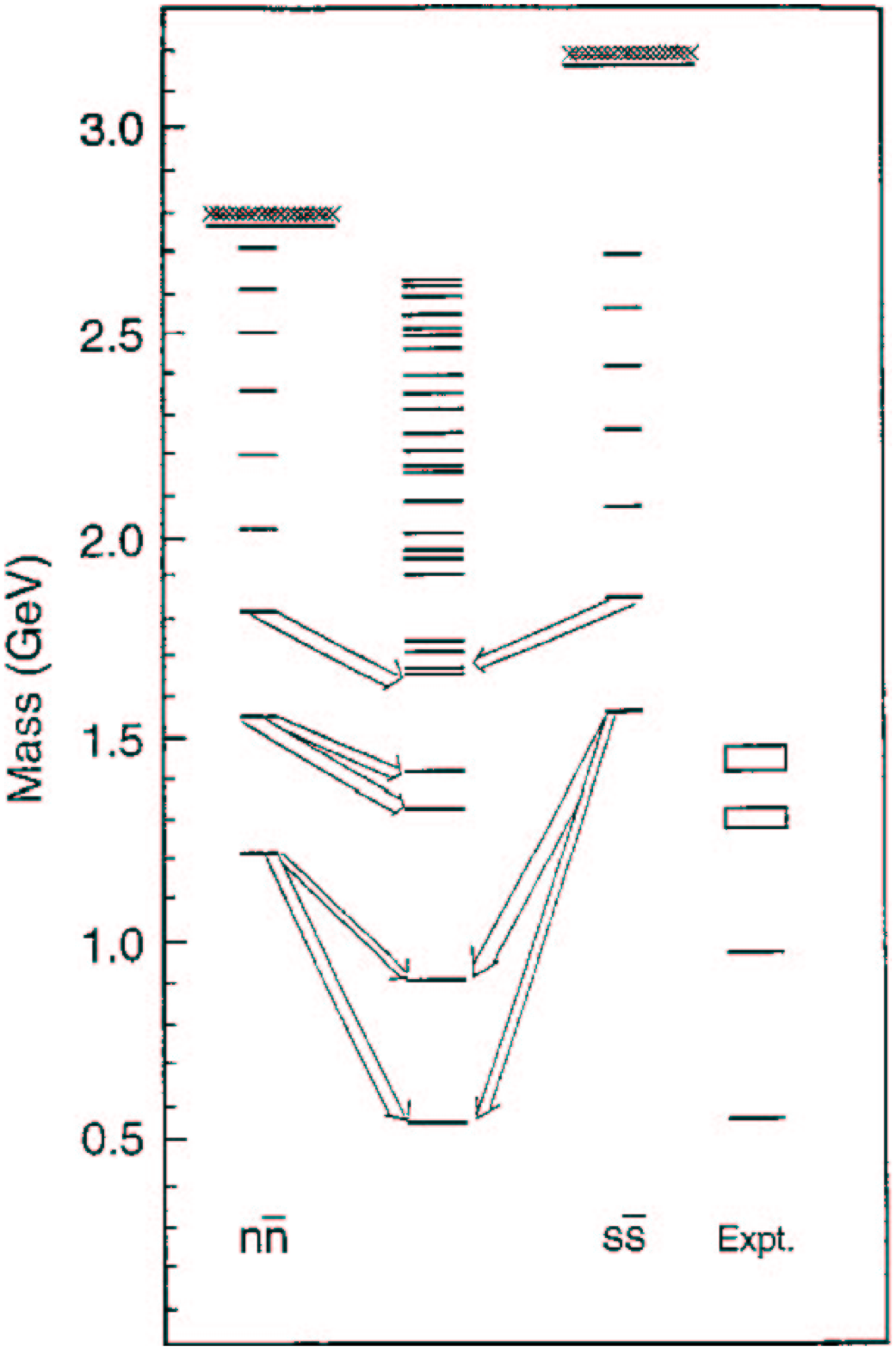}%
 %\caption{}
 \end{picture}

 \textbf{Fig. 4} Experimental and calculated spectra of the $\eta$ mesons. Columns 1 and 3 show the results for the confinement potential only.

\begin{center}
 %\begin{table}%[H] add [H] placement to break table across pages
% \begin{ruledtabular}
 \begin{tabular}{||@{\hspace{0.3cm}}
 c@{\hspace{0.3cm}}|@{\hspace{0.3cm}}c@{\hspace{0.3cm}}
 |@{\hspace{0.3cm}}c@{\hspace{0.3cm}}|@{\hspace{0.3cm}}c@{\hspace{0.3cm}}|@{\hspace{0.3cm}}
 c@{\hspace{0.3cm}}||@{\hspace{0.3cm}}
 c@{\hspace{0.3cm}}||}\hline\hline
                                             &Data                   &Set I          &Set II         &Set III        &Set IV\\\hline
 $m_\eta(547)$ [MeV]                         &$\cdots$               &538            &536            &527            &555  \\
 $m_{\eta^\prime}(958)$ [MeV]                &$\cdots$               &911            &942            &963            &949  \\
 $m_\eta(1295)$ [MeV]                        &$\cdots$               &1319           &1318           &1317           &1319 \\
 $m_\eta(1440)$ [MeV]                        &$\cdots$               &1414           &1416           &1419           &1411 \\\hline
 $\tilde f_\eta^{(8)}$ [MeV]                 &$\cdots$               &177.2          &178.6          &180.9          &163 \\
 $\tilde f_\eta^{(0)}$ [MeV]                 &$\cdots$               &27.59          &24.51          &18.95          &52.8 \\
 $\tilde f_{\eta^\prime}^{(8)}$ [MeV]        &$\cdots$               &-84.26         &-84.64         &-80.97         &-105 \\
 $\tilde f_{\eta^\prime}^{(0)}$ [MeV]        &$\cdots$               &159.2          &157.3          &156.0          &150 \\\hline
 $F_8$ [MeV]                                 &$(1.32\pm0.06)f_\pi)$  &               &               &               &    \\
                                             &$=174\pm8$ MeV         &179.3          &180.3          &181.9          &170 \\
 $F_0$ [MeV]                                 &$(1.37\pm0.07)f_\pi)$  &               &               &               &    \\
                                             &$=181\pm9$ MeV         &180.3          &178.2          &174.2          &190 \\\hline
 $\theta_\eta$                               &$(-5.7\pm2.7)^\circ$   &$-8.81^\circ$  &$-7.82^\circ$  &$-6.26^\circ$  &$-16.1^\circ$ \\
 $\theta_{\eta^\prime}$                      &$(-24.6\pm2.3)^\circ$  &$-28.0^\circ$  &$-28.0^\circ$  &$-26.4^\circ$  &$-38.2^\circ$ \\\hline
 $\theta_0$                                  &$(-7.0\pm2.7)^\circ$   &$-9.83^\circ$  &$-8.76^\circ$  &$-6.94^\circ$  &$-19.4^\circ$ \\
 $\theta_8$                                  &$(-21.5\pm2.4)^\circ$  &$-25.4^\circ$  &$-25.4^\circ$  &$-24.1^\circ$  &$-32.8^\circ$ \\
 $\theta_0-\theta_8$                         &$16.4^\circ$           &$15.6^\circ$   &$16.6^\circ$   &$17.2^\circ$   &$13.4^\circ$ \\\hline
 $\hat F_0$ [MeV]                            &$(1.21\pm0.07)f_\pi)$  &               &               &               &    \\
                                             &$=160\pm9$ MeV         &161            &159            &157            &158 \\
 $\hat F_8$ [MeV]                            &$=188\pm11$ MeV        &196            &198            &198            &194 \\\hline
 $G_D$ [\gev{-5}]                            &$\cdots$               &-180           &-200           &-220           &-161.6($G_{08}=0$) \\\hline\hline
% Lines of table here ending with \\
 \end{tabular}
 %\vspace{1.2cm}
% \end{ruledtabular}
% \end{table}
\end{center}
\newpage

\section{Chiral Symmetry Restoration and Deconfinement of Light Mesons at Finite Temperature}

\emph{Hu Li and C. M. Shakin , hep-ph/0209136}\\\\

Confinement model: \be V^C(r) = \kappa r \exp(-\mu_0r),
\,\,\,\,\,\,\,\,\,\,\,\,\,\,\,\,\,\,\,\,\,\,\,\,\,\,\,\,\,\,\,\,\,\,\,\,\,\,\,\,\,\,\,\,\,\,\,\,\,\,\,\,\,\,\,\,\,\,\,\,\,\,\,\,\,\,\,\,\,\,\,\ee
\be V^C(\vec k-\vec k\,^\prime)=-8\pi\kappa\left[\frac1{[(\vec
k-\vec k\,^\prime)^2+\mu^2]^2}-\frac{4\mu^2}{[(\vec k-\vec
k\,^\prime)^2+\mu^2]^3}\right]\,,\ee

\be V^C(\hat k-\hat k\,^\prime)=-8\pi\kappa\left[\frac1{[-(\hat
k-\hat k\,^\prime)^2+\mu^2]^2}-\frac{4\mu^2}{[-(\hat k-\hat
k\,^\prime)^2+\mu^2]^3}\right]\,,\ee

\be V^C(r,T)=\kappa r\exp[-\mu(T)r]\,,\ee \be
\mu(T)=\frac{\mu_0}{1-0.7(T/T_c)^2}\,,\ee

\be V_{max}(T)=\frac{\kappa[1-0.7(T/T_c)^2]}{\mu_0e}\,.\ee

\begin{picture}(200, 350)(0, -300)
 \centering\includegraphics[bb=0 0 250 350, angle=-90, scale=1.2]{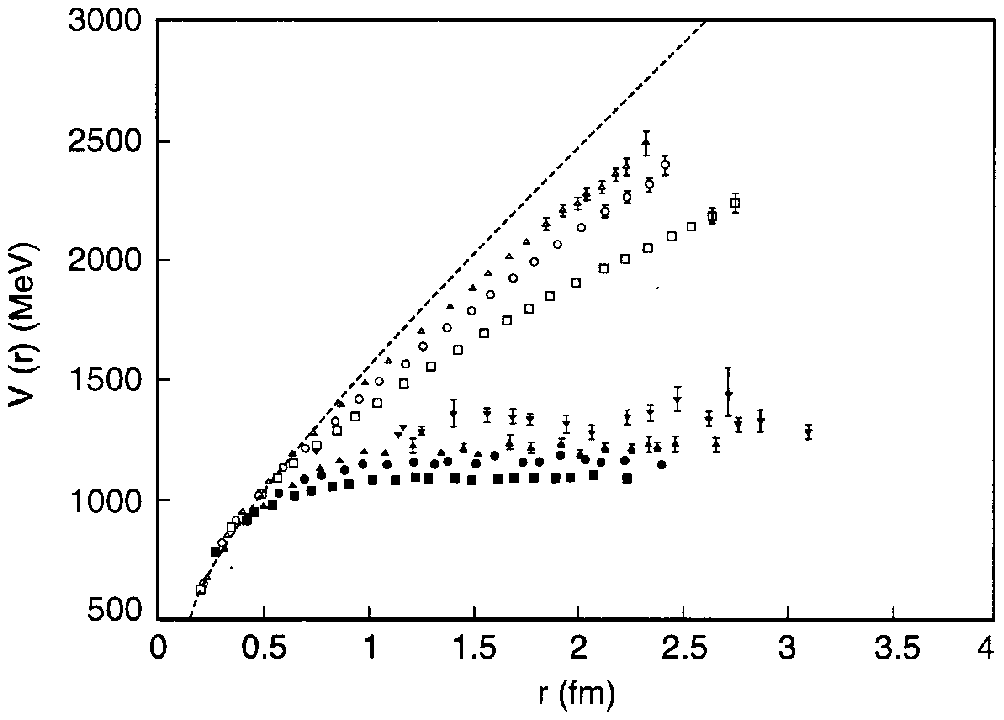}%
 %\caption{}
 \end{picture}

 \textbf{Fig. 5} A comparison of quenched (open symbols) and unquenched (filled symbols) results for
 the interquark potential at finite temperature. The dotted line is the zero temperature
 quenched potential. Here, the symbols for $T=0.80T_c$ [open triangle], $T=0.88T_c$
 [open circle], $T=0.94T_c$ [open square], represent the quenched
 results. The results with dynamical fermions are given at $T=0.68T_c$ [solid downward-pointing
 triangle], $T=0.80T_c$ [solid upward-pointing triangle], $T=0.88T_c$ [solid circle],
 and $T=0.94T_c$ [solid square].

\begin{picture}(300,200)
 \centering\includegraphics[bb=0 0 300 200, angle=0, scale=1.2]{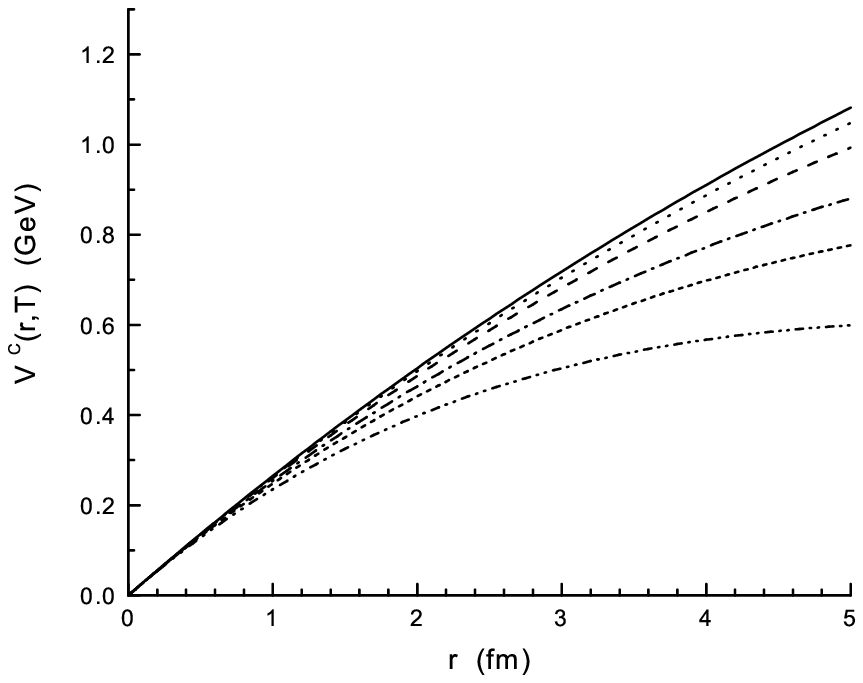}%
 %\caption{}
 \end{picture}

 \textbf{Fig. 6} The potential $V^C(r, T)$ is shown for $T/T_c=0$ [solid line],
 $T/T_c=0.4$ [dotted line], $T/T_c=0.6$ [dashed line], $T/T_c=0.8$ [dash-dot line],
 $T/T_c=0.9$ [short dashes], $T/T_c=1.0$ [dash-dot-dot line]. Here,
 $V^C(r,T)=\kappa r\exp[-\mu(T)r]$, with $\mu(T)=0.01\mbox{GeV}/[1-0.7(T/T_c)^2]$.

 \begin{picture}(300, 270)
 \centering\includegraphics[bb=0 0 300 200, angle=0, scale=1.2]{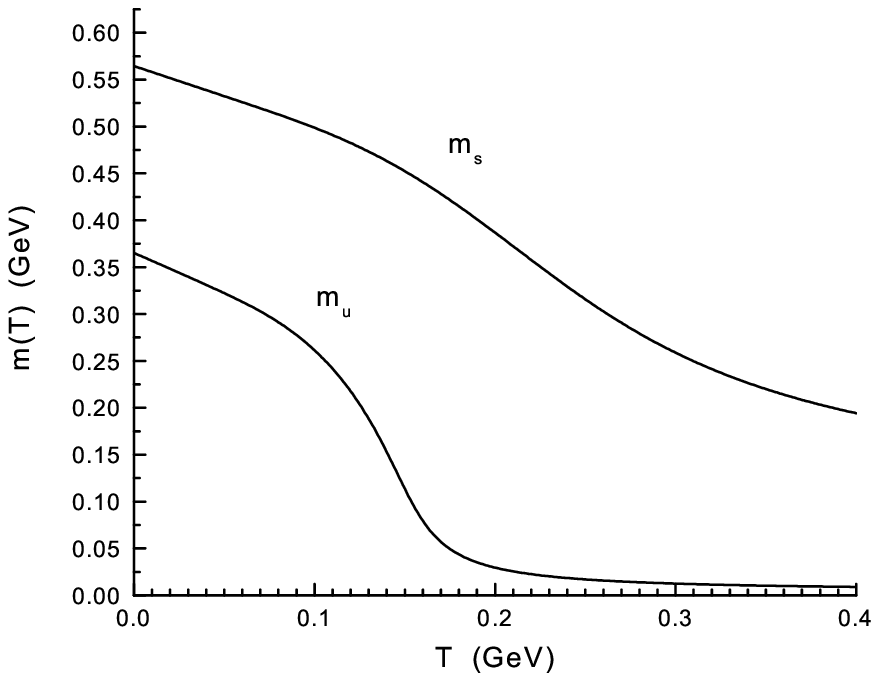}%
 %\caption{}
 \end{picture}

 \textbf{Fig. 7} Temperature dependent constituent mass values, $m_u(T)$ and $m_s(T)$,
 calculated using the equation below. are shown. Here $m_u^0=0.0055$ GeV,
 $m_s^0=0.120$ GeV, and $G(T)=5.691[1-0.17(T/T_c)]$, if we use
 Klevansky's notation.\\$m(T)=m^0+2G_S(T)N_C\frac{m(T)}{\pi^2}\int_0^\Lambda
dp\frac{p^2}{E_p}\tanh(\frac 1 2\beta E_p)$.

 \begin{picture}(280, 170)
 \centering\includegraphics[bb=0 0 280 235, angle=0, scale=1]{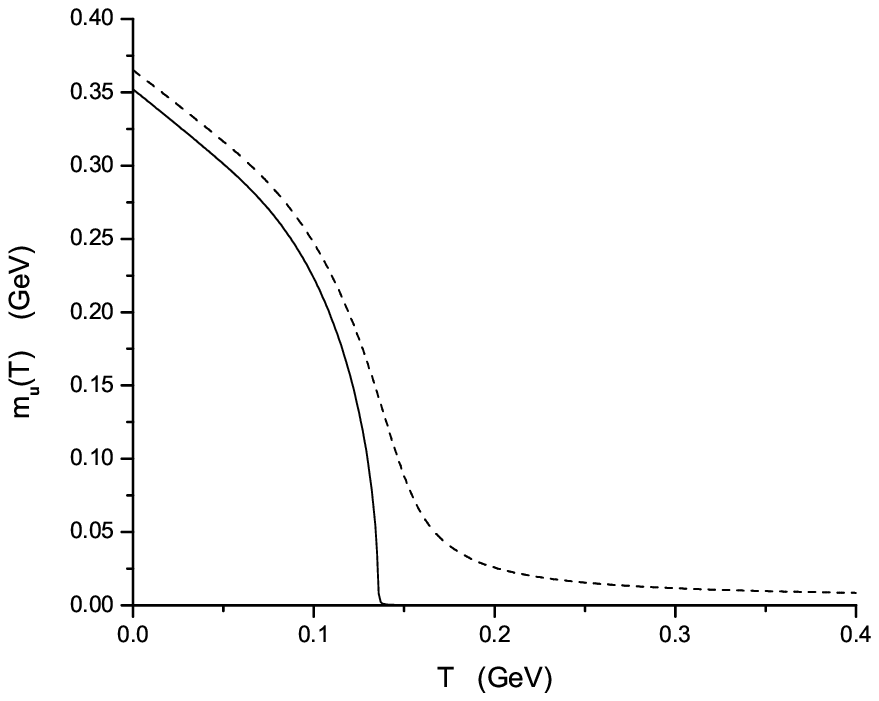}%
 %\caption{}
 \end{picture}

 \textbf{Fig. 8} Values of $m_u(T)$ are shown. The dashed curve is calculated with $m^0=5.50\,\mbox{MeV}$.
 Here, $G(T)=G\,[\,1-0.17\,(T/T_c)\,]$, with $G=5.691\,\mbox{GeV}^{-2}$ and $T_c=0.150\,\mbox{GeV}$. The
 solid curve is calculated with the same value of $G(T)$ and $T_c$, but with $m^0=0$. From the solid
 curve, we see that chiral symmetry is restored at $T=0.136\,\mbox{GeV}$ when $m^0=0$.

 \begin{picture}(440, 300)
 \centering\includegraphics[bb=0 0 440 670, angle=0, scale=0.4]{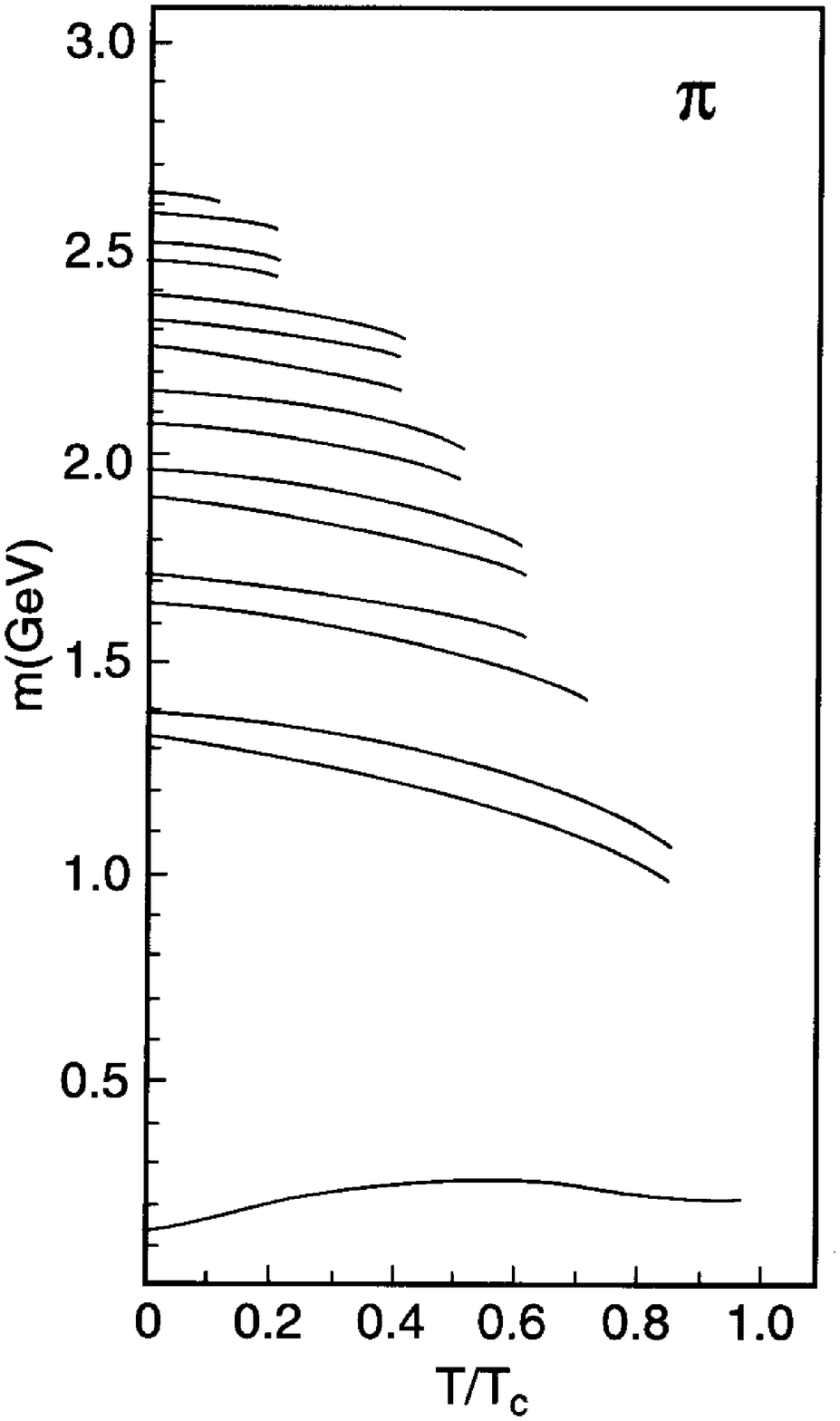}%
 %\caption{}
 \end{picture}

 \textbf{Fig. 9} The mass values of the pionic states calculated in this work
 with $G_\pi(T)=13.49[1-0.17\,T/T_c]$ GeV, $G_V(T)=11.46[1-0.17\,T/T_c]$ GeV. The value of the pion mass is
 0.223 GeV at $T/T_c=0.90$, where $m_u(T)=0.102$ GeV and $m_s(T)=0.449$ GeV. The
 pion is bound up to $T/T_c=0.94$, but is absent beyond that value.

\newpage
\section{Calculation of Hadronic Current Correlation
Functions at Finite Temperature}

\emph{Bing He, Hu Li, C. M. Shakin, and Qing Sun, Physical Review
C \textbf{67}, 065203 (2003)}\\\\

\begin{picture}(600, 120)
 \centering\includegraphics[bb=0 0 600 350, angle=0, scale=0.4]{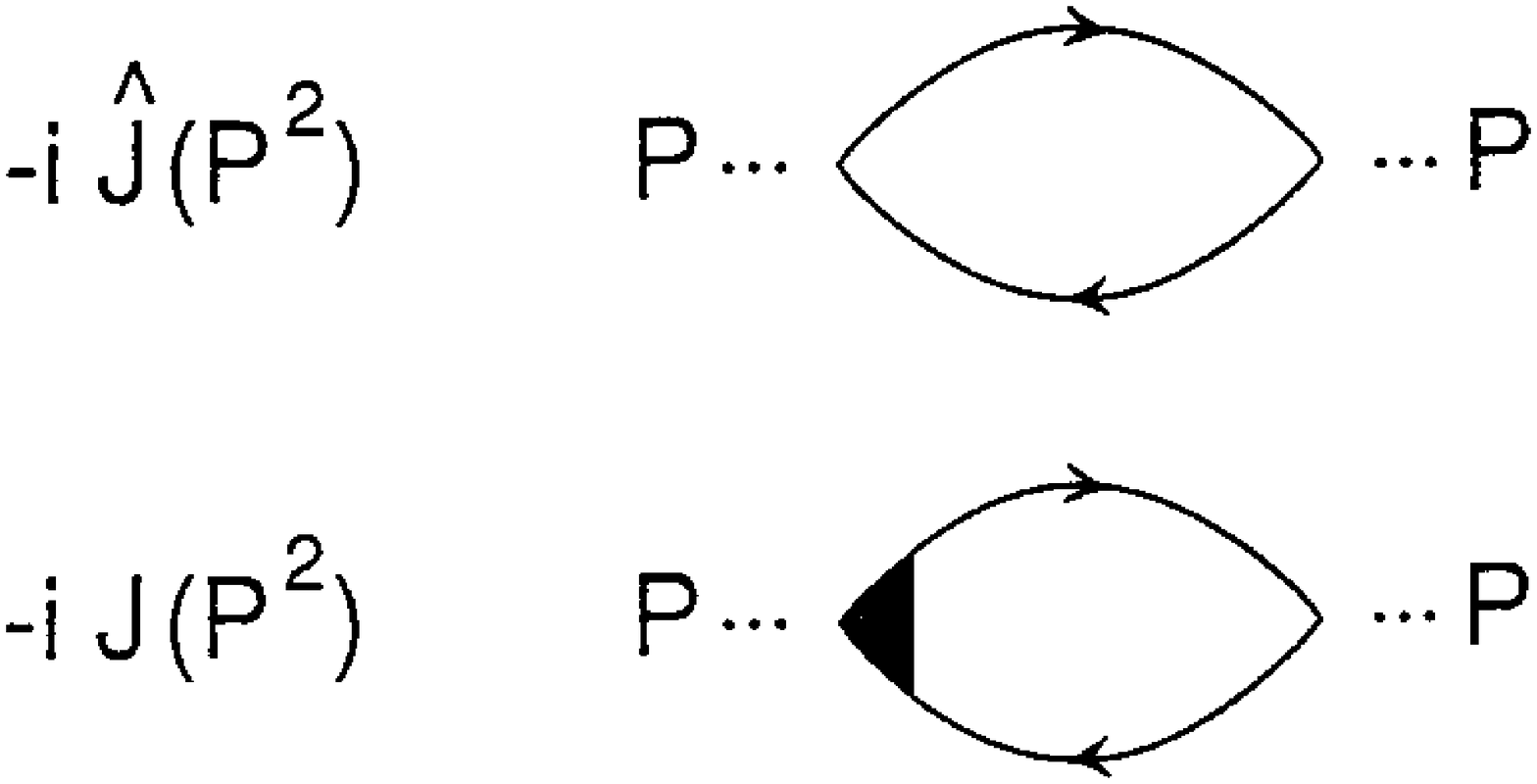}%
 %\caption{}
 \end{picture}

 \textbf{Fig. 10} The upper figure represents the basic polarization diagram of the
 NJL model in which the lines represent a constituent quark and a constituent
 antiquark. The lower figure shows a confinement vertex [filled triangular
 region] used in our earlier work. For the present work we neglect confinement
 for $T\geq1.2\,T_c$, with $T_c=150$ MeV.\\\\

For ease of reference, we present a discussion of our calculation
of hadronic current correlators. The procedure we adopt is based
upon the real-time finite-temperature formalism, in which the
imaginary part of the polarization function may be calculated.
Then, the real part of the function is obtained using a dispersion
relation. The result we need for this work has been already given
in the work of Kobes and Semenoff. (The quark momentum is $k$ and
the antiquark momentum is $k-P$. We will adopt that notation in
this section for ease of reference to the results presented in the
work of Kobes and Semenoff.) We write the imaginary part of the
scalar polarization function as \be
\label{e319.1}\mbox{Im}\,J_S(\textit{P}\,{}^2,
T)=\frac12(2N_c)\beta_S\,\epsilon(\textit{P}\,{}^0)\mytint
ke^{-\vec
k\,{}^2/\alpha^2}\left(\frac{2\pi}{2E_1(k)2E_2(k)}\right)\\\nonumber
\{(1-n_1(k)-n_2(k))
\delta(\textit{P}\,{}^0-E_1(k)-E_2(k))\\\nonumber-(n_1(k)-n_2(k))
\delta(\textit{P}\,{}^0+E_1(k)-E_2(k))\\\nonumber-(n_2(k)-n_1(k))
\delta(\textit{P}\,{}^0-E_1(k)+E_2(k))\\\nonumber-(1-n_1(k)-n_2(k))
\delta(\textit{P}\,{}^0+E_1(k)+E_2(k))\}\,.\ee Here,
$E_1(k)=[\,\vec k\,{}^2+m_1^2(T)\,]^{1/2}$. We have included a
Gaussian regulator, $\exp[\,-\vec k\,{}^2/\alpha^2\,]$, with
$\alpha=0.605$ GeV, which is the same as that used in most of our
applications of the NJL model in the calculation of meson
properties. We also note that \be n_1(k)=\frac1{e^{\,\beta
E_1(k)}+1}\,,\ee and \be n_2(k)=\frac1{e^{\,\beta E_2(k)}+1}\,.\ee
For the calculation of the imaginary part of the polarization
function, we may put $\ksq=m_1^2(T)$ and $(k-P)^2=m_2^2(T)$, since
in that calculation the quark and antiquark are on-mass-shell. In
Eq.\,(\ref{e319.1}) the factor $\beta_S$ arises from a trace
involving Dirac matrices, such that
\be \beta_S&=&-\mbox{Tr}[\,(\slr k+m_1)(\slr k-\slr P+m_2)\,]\\
&=&2P^2-2(m_1+m_2)^2\,,\ee where $m_1$ and $m_2$ depend upon
temperature. In the frame where $\vec P=0$, and in the case
$m_1=m_2$, we have $\beta_S=2P_0^2(1-{4m^2}/{P_0^2})$. For the
scalar case, with $m_1=m_2$, we find \be\label{e319.6}
\mbox{Im}\,J_S(P^2,
T)=\frac{N_cP_0^2}{4\pi}\left(1-\frac{4m^2(T)}{P_0^2}\right)^{3/2}
e^{-\vec k\,{}^2/\alpha^2}[\,1-2n_1(k)\,]\,,\ee where
\be\label{e319.7} \vec k\,{}^2=\frac{P_0^2}4-m^2(T)\,.\ee

For pseudoscalar mesons, we replace $\beta_S$ by
\be \beta_P&=&-\mbox{Tr}[\,i\gamma_5(\slr k+m_1)i\gamma_5(\slr k-\slr P+m_2)\,]\\
&=&2P^2-2(m_1-m_2)^2\,,\ee which for $m_1=m_2$ is $\beta_P=2P_0^2$
in the frame where $\vec P=0$. We find, for the $\pi$ mesons, \be
\mbox{Im}\,J_P(P^2,T)=\frac{N_cP_0^2}{4\pi}\left(1-\frac{4m^2(T)}{P_0^2}\right)^{1/2}
e^{-\vec k\,{}^2/\alpha^2}[\,1-2n_1(k)\,]\,,\ee where $ \vec
k\,{}^2={P_0^2}/4-m_u^2(T)$, as above. Thus, we see that, relative
to the scalar case, the phase space factor has an exponent of 1/2
corresponding to a \textit{s}-wave amplitude. For the scalars, the
exponent of the phase-space factor is 3/2.

For a study of vector mesons we consider \be
\beta_{\mu\nu}^V=\mbox{Tr}[\,\gamma_\mu(\slr k+m_1)\gamma_\nu(\slr
k-\slr P+m_2)\,]\,,\ee and calculate \be
g^{\mu\nu}\beta_{\mu\nu}^V=4[\,P^2-m_1^2-m_2^2+4m_1m_2\,]\,,\ee
which, in the equal-mass case, is equal to $4P_0^2+8m^2(T)$, when
$m_1=m_2$ and $\vec P=0$. This result will be needed when we
calculate the correlator of vector currents in the next section.
Note that, for the elevated temperatures considered in this work,
$m_u(T)=m_d(T)$ is quite small, so that $4P_0^2+8m_u^2(T)$ can be
approximated by $4P_0^2$, when we consider the vector current
correlation functions. In that case, we have \be
\mbox{Im}\,J_V(P^2,T) \simeq
\frac{2}{3}\mbox{Im}\,J_P(P^2,T)\,.\ee At this point it is useful
to define functions that do not contain that Gaussian regulator:
\be\mbox{Im}\,\tilde{J}_P(P^2,T)=\frac{N_cP_0^2}{4\pi}\left(1-\frac{4m^2(T)}{P_0^2}\right)^{1/2}[\,1-2n_1(k)\,]\,,\ee
and
\be\mbox{Im}\,\tilde{J}_V(P^2,T)=\frac{2}{3}\frac{N_cP_0^2}{4\pi}\left(1-\frac{4m^2(T)}{P_0^2}\right)^{1/2}[\,1-2n_1(k)\,]\,,\ee
We need to use a twice-subtracted dispersion relation to obtain
$\mbox{Re}\,\tilde{J}_P(P^2,T)$, or
$\mbox{Re}\,\tilde{J}_V(P^2,T)$. For example,
\be\label{e319.16}\mbox{Re}\,\tilde{J}_P(P^2,T)=\mbox{Re}\,\tilde{J}_P(0,T)+
\frac{P^2}{P_0^2}[\,\mbox{Re}\,\tilde{J}_P(P_0^2,T)-\mbox{Re}\,\tilde{J}_P(0,T)\,]+\\\nonumber
\frac{P^2(P^2-P_0^2)}{\pi}\int_{4m^2(T)}^{\tilde{\Lambda}^{2}}
ds\frac{\mbox{Im}\,\tilde{J}_P(s,T)}{s(P^2-s)(P_0^2-s)}\,,\ee
where $\tilde{\Lambda}^{2}$ can be quite large, since the integral
over the imaginary part of the polarization function is now
convergent. We may introduce $\tilde{J}_P(P^2,T)$ and
$\tilde{J}_V(P^2,T)$ as complex functions, since we now have both
the real and imaginary parts of these functions. We note that the
construction of either $\mbox{Re}\,J_P(P^2,T)$, or
$\mbox{Re}\,J_V(P^2,T)$, by means of a dispersion relation does
not require a subtraction. We use these functions to define the
complex functions $J_P(P^2,T)$ and $J_V(P^2,T)$.

In order to make use of Eq.\,(\ref{e319.16}), we need to specify
$\tilde{J}_P(0)$ and $\tilde{J}_P(P_0^2)$. We found it useful to
take $P_0^2=-1.0$ \gev2 and to put $\tilde{J}_P(0)=J_P(0)$ and
$\tilde{J}_P(P_0^2)=J_P(P_0^2)$. The quantities $\tilde{J}_V(0)$
and $\tilde{J}_V(P_0^2)$ are determined in an analogous function.
This procedure in which we fix the behavior of a function such as
$\mbox{Re}\tilde{J}_V(P^2)$ or $\mbox{Re}\tilde{J}_V(P^2)$ is
quite analogous to the procedure used in our earlier work. In that
work we made use of dispersion relations to construct a continuous
vector-isovector current correlation function which had the
correct perturbative behavior for large $P^2\rightarrow-\infty$
and also described that low-energy resonance present in the
correlator due to the excitation of the $\rho$ meson. In our
earlier work the NJL model was shown to provide a quite
satisfactory description of the low-energy resonant behavior of
the vector-isovector correlation function.

%\section{calculation of hadronic current correlation functions}
We now consider the calculation of temperature-dependent hadronic
current correlation functions. The general form of the correlator
is a transform of a time-ordered product of currents, \be iC(P^2,
T)=\int d^4xe^{iP\cdot x}<\!\!<T(j(x)j(0))>\!\!>\,,\ee where the
double bracket is a reminder that we are considering the finite
temperature case.

For the study of pseudoscalar states, we may consider currents of
the form $j_{P,i}(x)=\tilde{q}(x)i\gamma_5\lambda^iq(x)$, where,
in the case of the $\pi$ mesons, $i=1,2$ and $3$. For the study of
scalar-isoscalar mesons, we introduce
$j_{S,i}(x)=\tilde{q}(x)\lambda^i q(x)$, where $i=0$ for the
flavor-singlet current and $i=8$ for the flavor-octet current.

In the case of the pseudoscalar-isovector mesons, the correlator
may be expressed in terms of the basic vacuum polarization
function of the NJL model, $J_P(P^2, T)$. Thus, \be\label{e319.18}
C_P(P^2, T)=\tilde{J}_P(P^2, T)\frac{1}{1-G_{P}(T)J_P(P^2,
T)}\,,\ee where $G_P(T)$ is the coupling constant appropriate for
our study of $\pi$ mesons. We have found $G_P(T)=13.49$\gev{-2} by
fitting the pion mass in a calculation made at $T=0$, with $m_u =
m_d =0.364$ GeV. The result given in Eq.\,(\ref{e319.18}) is only
expected to be useful for small $P^2$, since the Gaussian
regulator strongly modifies the large $P^2$ behavior. Therefore,
we suggest that the following form is useful, if we are to
consider the larger values of $P^2$. \be\label{e319.19}
\frac{C_{P}(P^2, T)}{P^2}=\left[\frac{\tilde{J}_P(P^2,
T)}{P^2}\right] \frac{1}{1-G_P(T)J_P(P^2, T)}\,.\ee (As usual, we
put $\vec{P}=0$.) This form has two important features. At large
$P_0^2$, ${\mbox{Im}\,C_{P}(P_0, T)}/{P_0^2}$ is a constant, since
${\mbox{Im}\,\tilde{J}_{P}(P_0^2, T)}$ is proportional to $P_0^2$.
Further, the denominator of Eq.\,(\ref{e319.19}) goes to 1 for
large $P_0^2$. On the other hand, at small $P_0^2$, the
denominator is capable of describing resonant enhancement of the
correlation function. As we will see, the results obtained when
Eq.\,(\ref{e319.19}) is used appear quite satisfactory.

For a study of the vector-isovector correlators, we introduce
conserved vector currents $j_{\mu,
i}(x)=\tilde{q}(x)\gamma_{\mu}\lambda_i q(x)$ with i=1, 2 and 3.
In this case we define \be \label{e319.20}J_V^{\mu\nu}(P^2,
T)=\left(g\,{}^{\mu\nu}-\frac{P\,{}^\mu
P\,{}^\nu}{P^2}\right)J_V(P^2, T)\ee and \be\label{e319.21}
C_V^{\mu\nu}(P^2, T)=\left(g\,{}^{\mu\nu}-\frac{P\,{}^\mu
P\,{}^\nu}{P^2}\right)C_V(P^2, T)\,,\ee taking into account the
fact that the current $j_{\mu,\,i}(x)$ is conserved. (Note that
Eqs. (\ref{e319.20}) and (\ref{e319.21}) are valid for zero
temperature. However, we still use that form at finite temperature
for convenience.) We may then use the fact that \be J_V(P^2,T) =
\frac13g_{\mu\nu}J_V^{\mu\nu}(P^2,T)\ee and
\be\mbox{Im}\,J_V(P^2,T)&=&
\frac23\left[\frac{P_0^2+2m_u^2(T)}{4\pi}\right]
\left(1-\frac{4m_u^2(T)}{P_0^2}\right)^{1/2}e^{-\vec
k\,{}^2/\alpha^2}[\,1-2n_1(k)\,]\\
&\simeq& \frac{2}{3}\mbox{Im}J_P(P^2,T)\,.\ee (See
Eq.\,(\ref{e319.7}) for the specification of $k=|\vec k|$.) We
then have \be
C_V(P^2,T)=\tilde{J}_V(P^2,T)\frac1{1-G_V(T)J_V(P^2,T)}\,,\ee
where we have introduced \be\mbox{Im}\tilde{J}_V(P^2,T)&=&
\frac23\left[\frac{P_0^2+2m_u^2(T)}{4\pi}\right]
\left(1-\frac{4m_u^2(T)}{P_0^2}\right)^{1/2}[\,1-2n_1(k)\,]\\
&\simeq& \frac{2}{3}\mbox{Im}\tilde{J}_P(P^2,T)\,. \ee In the
literature, $\omega$ is used instead of $P_0$. We may define the
spectral functions \be\sigma_V(\omega,
T)=\frac{1}{\pi}\,\mbox{Im}\,C_V(\omega, T)\,,\ee and
\be\sigma_P(\omega, T)=\frac{1}{\pi}\,\mbox{Im}\,C_P(\omega,
T)\,,\ee

Since different conventions are used in the literature, we may use
the notation $\overline{\sigma}_P(\omega, T)$ and
$\overline{\sigma}_V(\omega, T)$ for the spectral functions given
there. We have the following relations: \be
\overline{\sigma}_P(\omega, T)=\sigma_P(\omega, T)\,,\ee and
\be\frac{\overline{\sigma}_V(\omega,
T)}{2}=\frac{3}{4}\sigma_V(\omega, T)\,,\ee where the factor 3/4
arises because there is a division by 4 in the literatures, while
we have divided by 3.

\section{Calculation of the Momentum Dependence of Hadronic
Current Correlation Functions at Finite Temperature}
\emph{Xiangdong Li, Hu Li, C. M. Shakin, Qing Sun and Huangsheng Wang, nucl-th/0405081}\\\\

We have calculated spectral functions associated with hadronic
current correlation functions for vector currents at finite
temperature. We made use of a model with chiral symmetry,
temperature-dependent coupling constants and temperature-dependent
momentum cutoff parameters. Our model has two parameters which are
used to fix the magnitude and position of the large peak seen in
the spectral functions. In our earlier work, good fits were
obtained for the spectral functions that were extracted from
lattice data by means of the maximum entropy method (MEM). In the
present work we extend our calculations and provide values for the
three-momentum dependence of the vector correlation function at
$T=1.5\,T_c$. These results are used to obtain the correlation
function in coordinate space, which is usually parametrized in
terms of a screening mass. Our results for the three-momentum
dependence of the spectral functions are similar to those found in
a recent lattice QCD calculation for charmonium [S. Datta, F.
Karsch, P. Petreczky and I. Wetzorke, hep-lat/0312037]. For a
limited range we find the exponential behavior in coordinate space
that is usually obtained for the spectral function for $T>T_c$ and
which allows for the definition of a screening mass.%\newpage

\begin{picture}(280, 250)
\centering\includegraphics[bb=0 0 280 235, angle=0, scale=1]{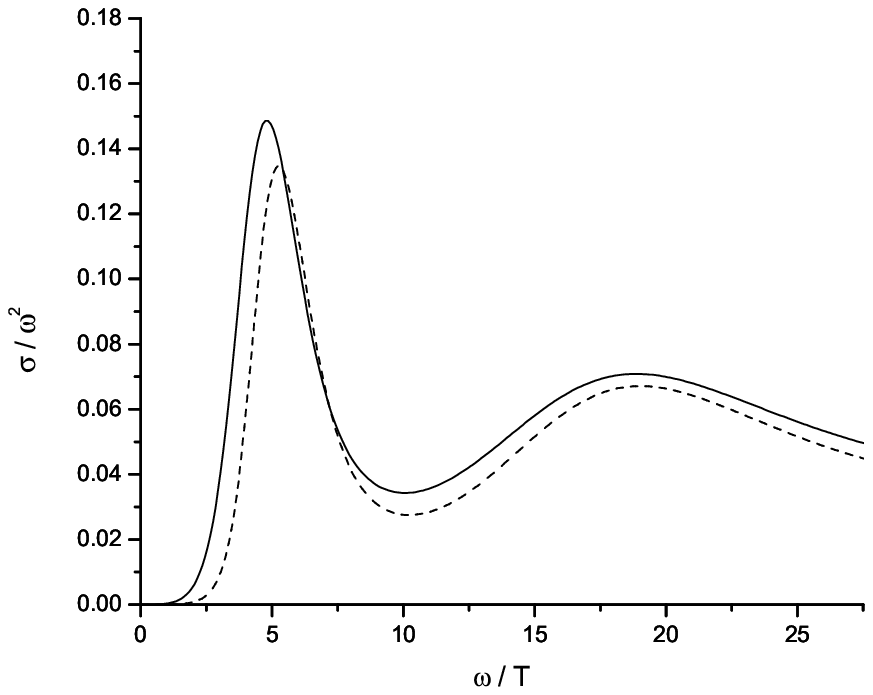}%
%\caption{}
%\label{f321a}
\end{picture}

\textbf{Fig. 11} The spectral functions $\sigma/\omega^2$ for
pseudoscalar states obtained by MEM are shown. The solid line is
for $T/T_c=1.5$ and the dashed line is for $T/T_c=3.0$. The second
peak is a lattice artifact.

\begin{picture}(280, 250)
\centering\includegraphics[bb=0 0 280 235, angle=0, scale=1]{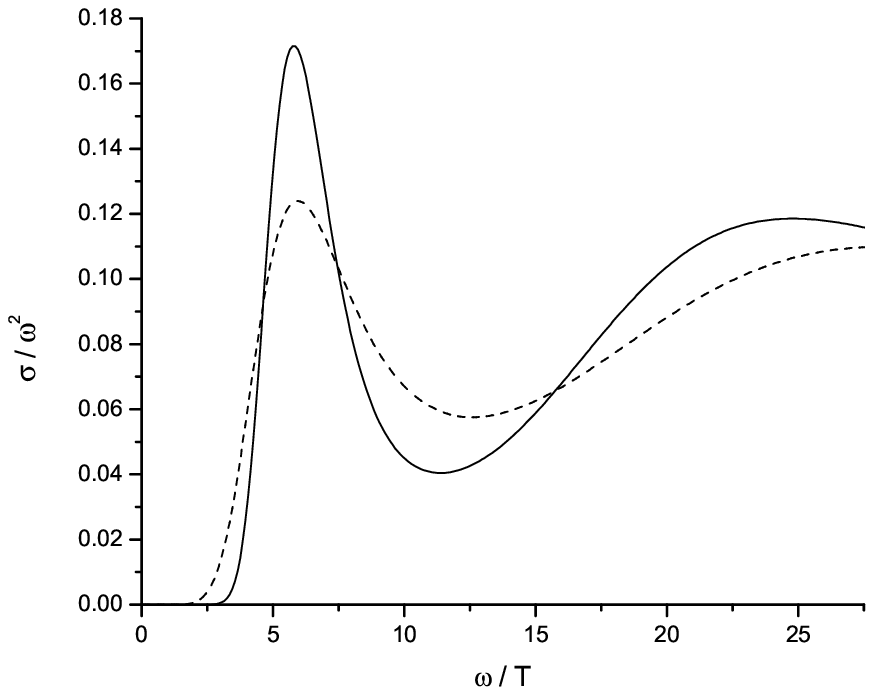}%
%\caption{}
\end{picture}

\textbf{Fig. 12} The spectral functions $\sigma/\omega^2$ for
vector states obtained by MEM are shown. The second peak is a
lattice artifact.

\begin{picture}(280, 200)
\centering\includegraphics[bb=0 0 280 235, angle=0, scale=1]{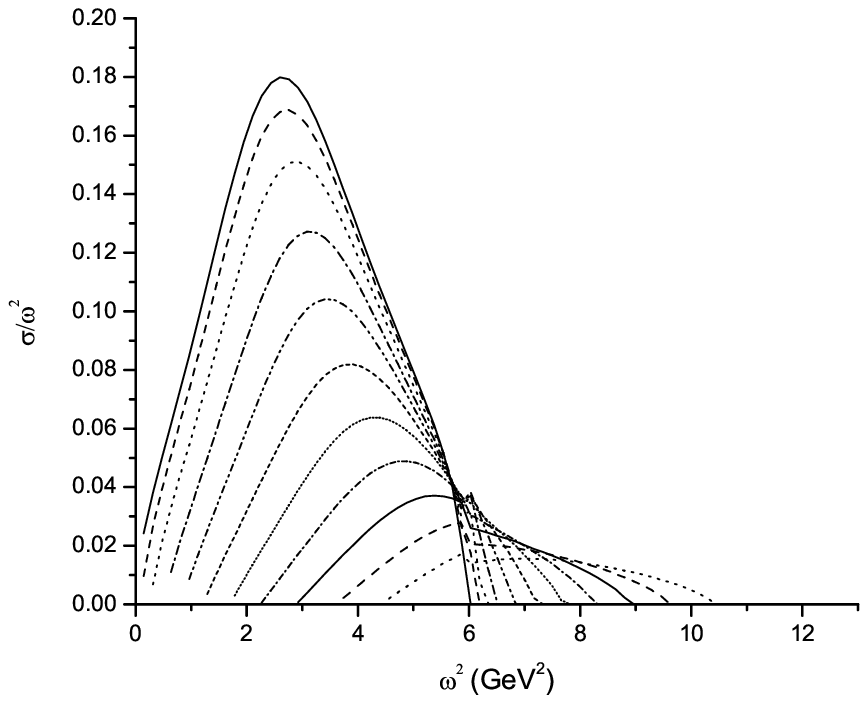}%
%\caption{}
\end{picture}

\textbf{Fig. 13} The imaginary part of the correlator
$\sigma(\omega)/\omega^2$ is shown for various values of $|\vec
P|$ as a function of $\omega^{2}$. Starting with the topmost curve
the values of $|\vec P|$ in GeV units are 0.10, 0.30, 0.50, 0.70,
0.90, 1.10, 1.30, 1.50, 1.70, 1.90 and 2.10. Here we have used
$G_S=1.2$ GeV$^{-2}$ and $k_{max}=1.22$ GeV.

\begin{picture}(280, 260)
\centering\includegraphics[bb=0 0 280 235, angle=0, scale=1]{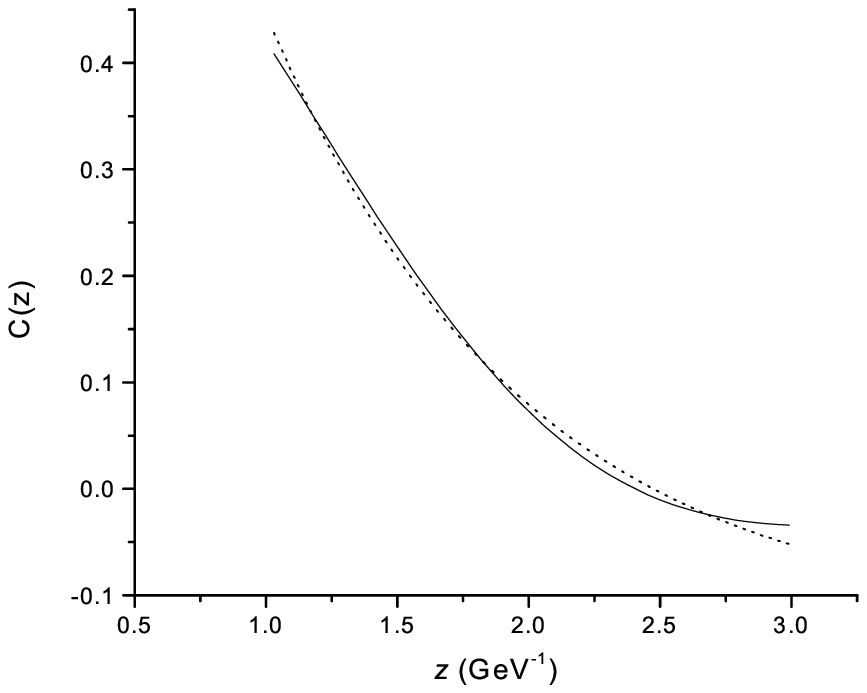}%
%\caption{}
\end{picture}

\textbf{Fig. 14} The correlation function $C(z)$ is shown. The
dotted line represents a fit using an exponential function.\\
Here, $
C(z)=\frac12\int_{-\infty}^\infty\,dP_ze^{iP_zz}\int_0^\infty\,d\omega\frac{\sigma(\omega,
0, 0, P_z)}\omega\,. $ We may also use the form $
C(z)=\frac14\int_{-\infty}^\infty\,dP_ze^{iP_zz}\int_0^\infty\,dP^2\,\frac{\sigma(P^2,
0, 0, P_z)}{P^2}\,. $
\newpage

\section{Quark propagation in the quark-gluon plasma}

\emph{Xiangdong Li, Hu Li, C. M. Shakin and Qing Sun, Physical
Review C, \textbf{69},
065201 (2004)}\\\\

It has recently been suggested that the quark-gluon plasma formed
in heavy-ion collisions behaves as a nearly ideal fluid. That
behavior may be understood if the quark and antiquark mean-free-
paths are very small in the system, leading to a ``sticky
molasses" description of the plasma, as advocated by the Stony
Brook group. This behavior may be traced to the fact that there
are relatively low-energy $q\overline{q}$ resonance states in the
plasma leading to very large scattering lengths for the quarks.
These resonances have been found in lattice simulation of QCD
using the maximum entropy method (MEM). We have used a chiral
quark model, which provides a simple representation of effects due
to instanton dynamics, to study the resonances obtained using the
MEM scheme. In the present work we use our model to study the
optical potential of a quark in the quark-gluon plasma and
calculate the quark mean-free-path. Our results represent a
specific example of the dynamics of the plasma as described by the
Stony Brook group.
%\newpage

\begin{picture}(280,250)
\centering\includegraphics[bb=0 0 280 235, angle=0, scale=1]{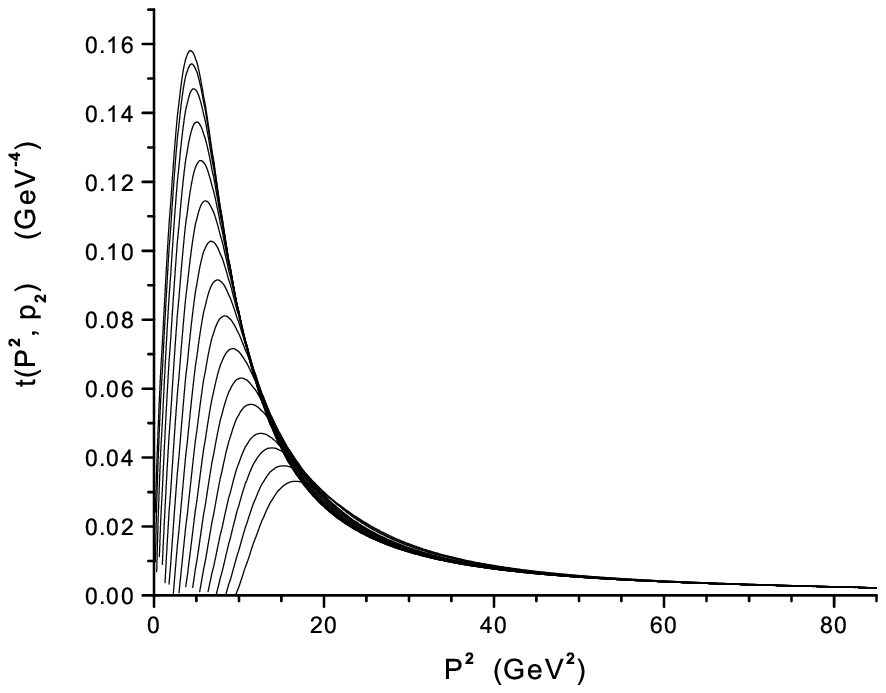}%
%\caption{}
\end{picture}

\textbf{Fig. 15} Values of $t(P^2, p_2)$ are shown for various
values of the quark momentum $|\vec{p}_2|$. Starting with the
uppermost curve, the $|\vec{p}_2|$ values in GeV units are 0.01,
0.03, 0.05, 0.07, 0.09, 0.11, 0.13, 0.15, 0.17, 0.19, 0.21, 0.23,
0.25, 0.27, 0.29 and 0.31. (For large $P^2$, we have $t(P^2,
p_2)\simeq(1/\pi P^2)G$.) Here $P^2=(p_1+p_2)^2$, where $p_1$ is
the antiquark momentum.

\begin{picture}(280, 240)
\centering\includegraphics[bb=0 0 280 235, angle=0, scale=1]{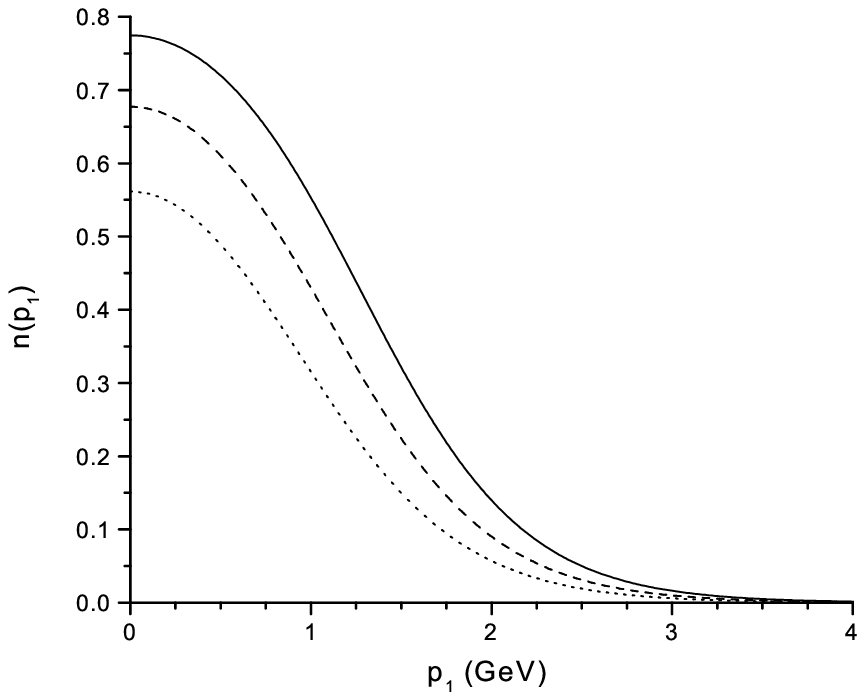}%
%\caption{}
\end{picture}

\textbf{Fig. 16} Values of $n(p_1)$ are shown for $\mu=1.1$\,GeV
(dotted curve), $\mu=1.3$\,GeV (dashed curve) and $\mu=1.5$\,GeV
(solid curve). Here $T=1.5\,T_c$ with $T_c=270$\,MeV.

\begin{picture}(280,260)
\centering\includegraphics[bb=0 0 280 235, angle=0, scale=1]{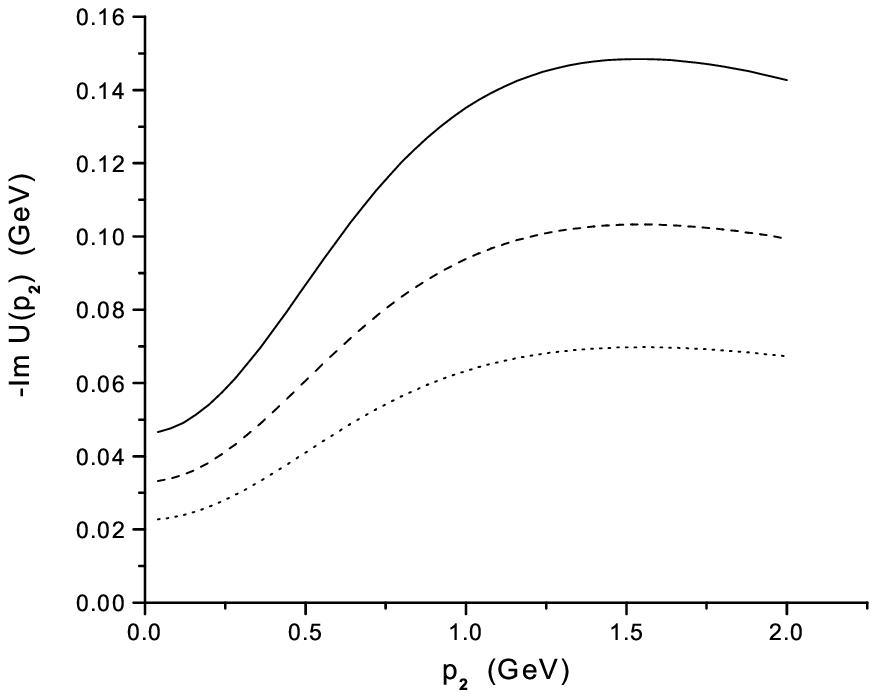}%
%\caption{}
\end{picture}

\textbf{Fig. 17} The imaginary part of the quark optical potential
is shown for $\mu=1.1$\,GeV (dotted curve), $\mu=1.3$\,GeV (dashed
curve) and $\mu=1.5$\,GeV (solid curve). (We recall that the
nucleon-nucleus imaginary optical potential is about 0.01\,GeV in
magnitude.)

\newpage

\section{Calculation of Screening Masses in a Chiral Quark Model}

\emph{Xiangdong Li, Hu Li, C. M. Shakin and Qing Sun, nucl-th/0405035}\\\\

We consider a simple model for the coordinate-space vacuum
polarization function which is often parametrized in terms of a
screening mass. We discuss the circumstances in which the value
$m_{sc}=\pi T$ is obtained for the screening mass. In the model
considered here, that result is obtained when the momenta in the
relevant vacuum polarization integral are small with respect to
the first Matsubara frequency.

In order to present our results in the simplest form, we consider
only the scalar interaction proportional to $(\overline{q}q)^{2}$.
We also extend the definition of $\sigma(\omega,T)$ to include a
dependence upon the total moment of the quark and antiquark
appearing in the polarization integral. Thus we consider the
imaginary part of the correlator,
$\sigma(\omega,\overrightarrow{P})$. Since we place
$\overrightarrow P$ along the \emph{z}-axis this quantity may be
written as $\sigma(\omega, 0, 0, P_z)$. In this work we will
present our result for the coordinate-dependent correlator $C(z)$
which is proportional to the correlator, \be
C(z)=\frac12\int_{-\infty}^\infty\,dP_ze^{iP_zz}\int_0^\infty\,d\omega\frac{\sigma(\omega,
0, 0, P_z)}\omega\,. \ee We may also use the form \be
C(z)=\frac14\int_{-\infty}^\infty\,dP_ze^{iP_zz}\int_0^\infty\,dP^2\,\frac{\sigma(P^2,
0, 0, P_z)}{P^2}\,. \ee

We have made a study of the screening mass in a simple model in
order to understand the origin of exponential behavior for the
correlator. We consider the Matsubara formalism and note that the
quark propagator may be written, with $\beta=1/T$, \be
S_\beta(\overrightarrow{k},\omega_n)=\frac{\gamma^0(2n+1)\pi/\beta+
\overrightarrow{\gamma}\cdot\overrightarrow{k}-M}{(2n+1)^2\pi^2/\beta^2+\overrightarrow{k}^2+M^2}\,.\ee
For bosons the vacuum polarization function  is given as,
\be\Pi(\overrightarrow{p},p^0)=\frac{g^2}{2\beta}\sum_n\frac{d^3k}{(2\pi)^3}\,
\frac1{\dfrac{4n^2\pi^2}{\beta^2}+\overrightarrow{k}^2+M^2}\cdot\frac{1}
{\left(\dfrac{2n\pi}{\beta}+p^0\right)^2+(\overrightarrow{k}+\overrightarrow{p})^2+M^2}\,.\ee

We modify the last equation to refer to fermions. In this case the
Matsubara frequencies are \be\omega_n=\frac{(2n+1)\pi}\beta\ee and
we have
\be\Pi(\overrightarrow{p},p^0)=\frac{g^2}{2\beta}\mbox{Tr}\mytint
k\frac{\left[\left(\gamma^0\pi/\beta+\overrightarrow{\gamma}\cdot\overrightarrow{k}\right)
\left(\gamma^0(p^0+\pi/\beta)+\overrightarrow{\gamma}\cdot(\overrightarrow{k}+\overrightarrow{p})\right)\right]}
{\left(\dfrac{\pi^2}{\beta^2}+\overrightarrow{k}^2\right)\left[\left(\dfrac\pi\beta+p^0\right)^2+
(\overrightarrow{k}+\overrightarrow{p})^2\right]}\,,\ee if we keep
only the first term in the sum, where $\omega_0=\pi/\beta$. As a
next step we drop $p^0$, so that we have
\be\Pi(\overrightarrow{p},0)=\frac{g^2}{2\beta}\mbox{Tr}\mytint
k\frac{\left[\left(\gamma^0\pi/\beta+\overrightarrow{\gamma}\cdot\overrightarrow{k}\right)
\left(\gamma^0\pi/\beta+\overrightarrow{\gamma}\cdot(\overrightarrow{k}+\overrightarrow{p})\right)\right]}
{\left(\left(\dfrac{\pi}{\beta}\right)^2+\overrightarrow{k}^2\right)\left[\left(\dfrac\pi\beta\right)^2+
(\overrightarrow{k}+\overrightarrow{p})^2\right]}\,.\ee We then
take $\overrightarrow{p}$ along the $z$ axis and write
$\Pi(p_z)=\Pi(\overrightarrow{p},0)$. We define \be C(z)=\int
dp_z\,e^{ip_zz}\,\Pi(p_z)\,.\ee In our calculation we replace
$g^2/2\beta$ by unity and use a sharp cutoff so that
$|\overrightarrow{k}|<k_{max}$.
%\newpage
\begin{picture}(280,250)
\includegraphics[bb=0 0 280 235, angle=0, scale=1]{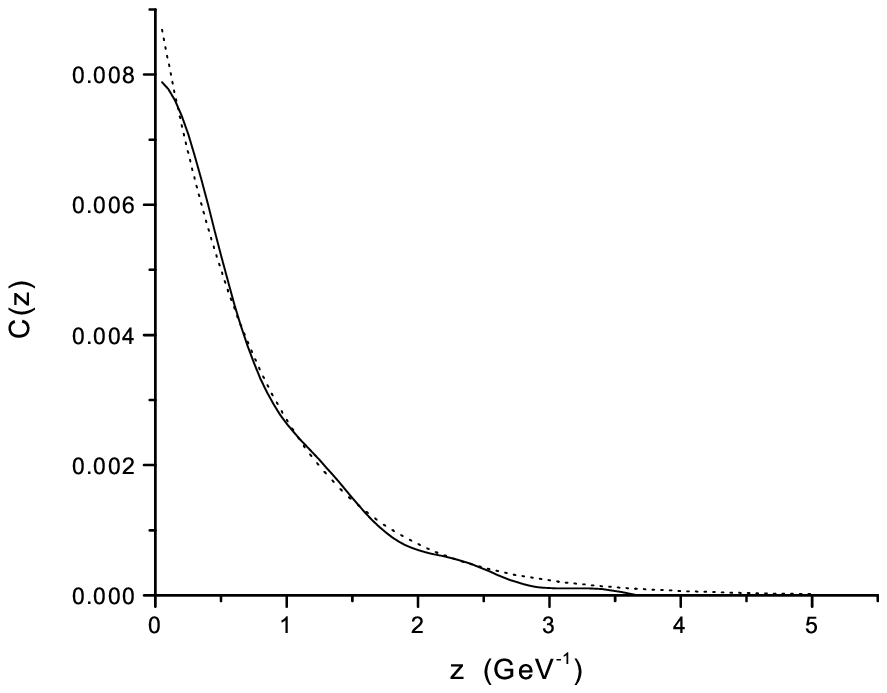}%
%\caption{}
\end{picture}

\textbf{Fig. 18} The function $C(z)$ is shown for a sharp cutoff
of $k_{max}=0.1$ GeV. The dotted line represents an exponential
fit to the curve using $m_{sc}=1.23$ GeV. (We recall that $\pi T$
is equal to 1.27 GeV.)

\begin{picture}(280,250)
\includegraphics[bb=0 0 280 235, angle=0, scale=1]{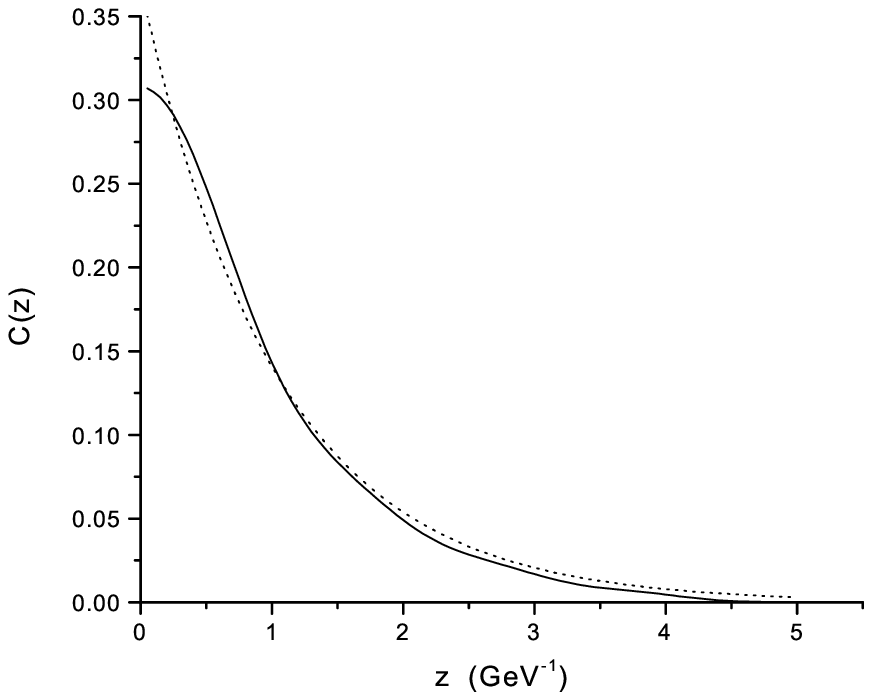}%
%\caption{}
\end{picture}

\textbf{Fig. 19} The function $C(z)$ of is shown for a sharp
cutoff of $k_{max}=0.4$ GeV. The dotted line represents an
exponential fit to the curve using $m_{sc}=0.961$ GeV. (We recall
that $\pi T$ is equal to 1.27 GeV.)

\end{document}